# RR Lyrae stars in eclipsing systems – historical candidates

J. Liška [1,2,3], M. Skarka [2,4], P. Hájková [5], R. F. Auer [2,6]

(1) Department of Theoretical Physics and Astrophysics, Faculty of Science, Masaryk University, Kotlářská 2, 611 37 Brno, Czech Republic, jiriliska@post.cz
(2) Variable Star and Exoplanet Section of the Czech Astronomical Society, Observatory Valašské Meziříčí, Vsetínská 78, 757 01 Valašské Meziříčí, Czech Republic
(3) Brno Observatory and Planetarium, Kraví hora 2, 616 00 Brno, Czech Republic
(4) Konkoly Observatory, Research Centre for Astronomy and Earth Sciences, Hungarian Academy of Sciences, H-1121 Budapest, Konkoly Thege Miklós út. 15-17, Hungary, marek.skarka@csfk.mta.hu
(5) Gymnázium Brno-Řečkovice, Terezy Novákové 2, 621 00 Brno, Czech Republic, petra-hajkova@centrum.cz
(6) South-Moravian-Observatory, Chudčice 273, 664 71 Veverská Bítýška, Czech Republic, auer.reinhold@gmail.com

**Abstract:** Discovery of binary systems among RR Lyrae stars belongs to challenges of present astronomy. So far, none of classical RR Lyrae stars was clearly confirmed that it is a part of an eclipsing system. For this reason we studied two RR Lyrae stars, VX Her and RW Ari, in which changes assigned to eclipses were detected in sixties and seventies of the 20[th] century. In this paper our preliminary results based on analysis of new photometric measurements are presented as well as the results from the detailed analysis of original measurements. A new possible eclipsing system, RZ Cet was identified in the archive data. Our analysis rather indicates errors in measurements and reductions of the old data than real changes for all three stars.

**Abstrakt:** Objevení dvojhvězd mezi RR Lyrae hvězdami patří k výzvám současné astronomie. Zatím u žádné klasické RR Lyrae hvězdy nebylo prokazatelně potvrzeno, že je součástí zákrytového systému. Studovali jsme proto dvě RR Lyrae hvězdy, VX Her a RW Ari, u kterých byly v 60. a 70. letech 20. století detekovány změny přisouzené zákrytům. V tomto příspěvku jsou prezentovány předběžné výsledky založené na nových fotometrických měřeních a výsledky z detailního rozboru původních měření. V archivních datech se podařilo objevit nový možný zákrytový systém RZ Cet. Naše analýza poukazuje spíše na chyby při měření a zpracování u starších dat, než na reálné změny u všech tří hvězd.

## Introduction

Detection of eclipsing binary (EB) systems with RR Lyrae component has the highest priority in the beginning of the 21[st] century. According to the literature and new database of candidates for RR Lyrae stars in binary systems (RRLyrBinCan, Liška et al. 2015, Liška & Skarka 2016) several EB candidates were identified in the past. Many of known candidates were discovered accidentally in photometric data as strange drop(s) in the light curve – VX Her (Fitch et al. 1966), RW Ari (Wiśniewski 1971), V80 in Ursae Minor Dwarf galaxy (Kholopov 1971), and RZ Cet (Liška, 2015).

Systematic search for this kind of EB systems is still limited. Richmond (2011), who analysed data for RR Lyrae stars from MACHO database (more than three thousand objects), did not succeed. On the other hand, research based on similar data quality from OGLE project, gave promising results. Soszyński et al. 2003 found 3 variable stars in the Large Magellanic Cloud (LMC) which show pulsating and eclipsing behaviour. Unfortunately these stars were probably only blends of EB system with RR Lyrae star (Soszyński et al. 2003, Prša et al. 2008). Soszyński et al. 2009 found another object in LMC OGLE-LMC-03541 which could also be blend. Soszyński et al. 2011 detected brightness drops in light curves of 3 RR Lyrae stars from the Galactic Bulge which could be eclipses, but expected orbital periods are unknown. They also revealed the best studied candidate, OGLE-BLG-02792, which, unfortunately, does not contain a classical RR Lyrae component. This star became a prototype of a new class of pulsating stars – Binary Evolutionary Pulsators (Pietrzyński et al. 2012, Smolec et al. 2013).

The number of candidates is still very low regarding tens of thousands known RR Lyrae stars. Therefore, search for new objects and confirmation of all known candidates is very desirable. In this study we present an investigation of measurements of the EB candidates which were revealed on the historical photometric data (VX Her, RW Ari, RZ Cet) because detailed analysis of these objects is missing.

## Observations and analysis

We used original photoelectric *UBV* measurements for VX Her from Fitch et al. (1966) and for RW Ari and RZ Cet from Bookmeyer et al. (1977). In addition, we used data from ASAS-3 database (Pojmanski 2002) for VX Her and RZ Cet. In RZ Cet also data from Sturch (1966) and Epstein (1969) was investigated. To complete available data we obtained new CCD observations of RW Ari and VX Her at Masaryk University Observatory (MUO) and Brno Observatory and Planetarium, both in Brno, Czech Republic. Old photoelectric data was compared with the newest CCD measurements. Phased light curves were reconstructed for all stars and phase shifts caused by period changes were reduced. More details are in Liška et al. (2015) or will be present in forthcoming paper.

## VX Her

Variability of VX Her was discovered by Raymond (see Campbell & Pickering 1917) based on photographic measurements, who determined pulsation period of 0.365 d which is 1-day alias of the correct period. The period was refined by Esch (1918) to 0.456 d.

Fitch et al. (1966) obtained photoelectric photometry in *UBV* filters for VX Her in 5 nights. The brightness of the star was anomalous in one of their observing nights. It was about 0.7 mag fainter (in *V*-band) than in normal minimum and its brightness was almost constant. They proposed eclipsing binary hypothesis for explanation of the observed brightness depression. Their study is probable the first publication in history, where eclipsing behaviour among RR Lyrae stars is mentioned. It is quite surprising that for a long time no other study was focused on binarity of VX Her. Until now, probably only Perry et al. (2015) tried to detect eclipses in a systematic way, but their dataset is not sufficiently homogeneous and extensive. Our attempt to test binary hypothesis using analysis of *O–C* diagram is described in Liška et al. (2015).

For this study we performed detailed analysis of data from Fitch et al. (1966) and compared it with our results based on MUO and ASAS-3 (*V*-band) data. We found that in the strange night (JD 2439217) the star had constant brightness despite the fact that is should increase its brightness towards maximum (Fig. 1, the left panel). Colour changes differ of about 0.1 mag in *B–V* and *U–B* than in normal nights (Fig. 1, the right panel). It can correspond to different colour of the secondary component, if the binary hypothesis is valid. Nevertheless, these distinctions can be explained also by fault identification of measured star in the night JD 2439217. For example, there are 3 stars in a close vicinity of VX Her (closer than 15') which have similar brightness and colour indices (see Table 1). The most similar is TYC 1510-155-1. In addition, our photometric data shows no variation larger than 0.02 mag.

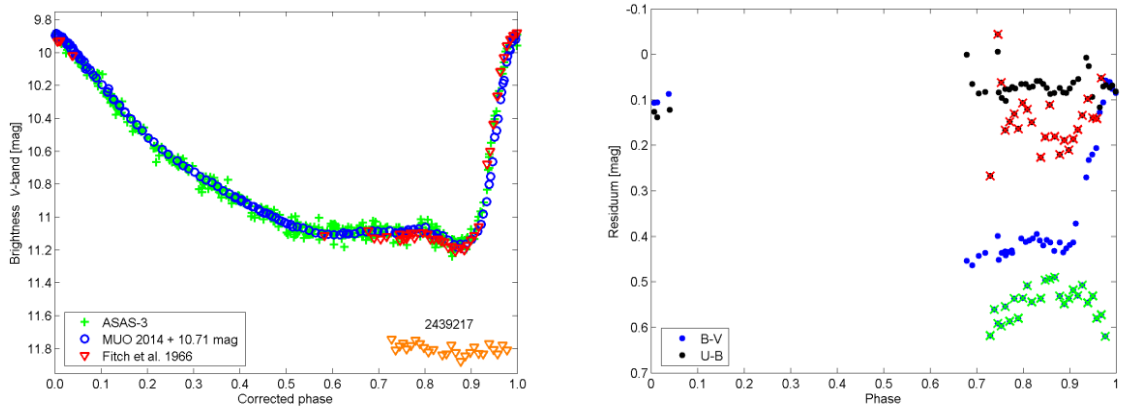

**Figure 1:** The light curve of VX Her in *V*-band from Fitch et al. (1966), ASAS-3 database and our observations phased with pulsation period (the left panel). Variations in colours in observations from Fitch et al. (1966) – green and red crosses mark *B−V* and *U−B* indices from the night JD 2439217.

**Table 1.** Stars in the vicinity of VX Her (up to 15') with similar brightness as VX Her (in JD 2439217). Coordinates and magnitudes were taken from Simbad database (Wenger et al. 2000).

| Star | Angular distance ['] | RA [$^h$ $^m$ $^s$] | DEC [° ' ''] | *V* [mag] | *B* [mag] |
|---|---|---|---|---|---|
| TYC 1510-149-1 | 3.2 | 16 30 27.66 | +18 22 46.7 | 11.74(13) | 12.65(21) |
| TYC 1510-36-1 | 7.8 | 16 30 20.51 | +18 28 09.2 | 11.93(15) | 12.13(12) |
| TYC 1510-155-1 | 11.2 | 16 30 05.56 | +18 29 29.9 | 11.82(13) | 12.31(15) |
| VX Her (JD 2439217) | – | 16 30 40.80 | +18 22 00.5 | 11.80(3) | 12.35(3) |

## RW Ari

RRc type star RW Ari has a similar history like VX Her. Detre (1937) found its variability on photographic plates and established incorrect period of 0.2614151 d (1-day alias). Notni (1962) determined correct value of 0.3543184 d.

For our goal the most important study is the paper by Wiśniewski (1971), who obtained observations of RW Ari in 19 nights (measurements were published later in Bookmeyer et al. 1977). Behaviour of the star was anomalous in three nights, in JD 2439384 and 2439505 it was fainter of about 0.6 mag (in $V$-band), and in 2439411 only of about 0.1 mag than in normal state. He proposed that the variation is caused by eclipses by another object in the system, and the deeper and shallower drops are primary and secondary minima, respectively. He estimated the orbital period as 3.1754 d. According to the colour changes during primary eclipse (in JD 2439384 $B$–$V$ was growing with decreasing $V$-brightness) Wiśniewski estimated that the secondary component is bluer and hotter than RW Ari.

Several authors tried to confirm EB hypothesis of RW Ari. Woodward (1972) and Sidorov (1978) independently found brightness depression of about 0.2 mag in one night in photographic data from Detre (1937). On the other hand, Penston (1972), Edwards (1978), Goranskij & Shugarov (1979) detected nothing unusual in their new observations and their results do not support the binarity of RW Ari. Another strange finding comes from Dahm (1992) and his new analysis of data from Wiśniewski (published in Bookmeyer et al. 1977). Dahm found more atypical variations in another three nights (depressions and extreme brightening, see his Figs. 1 and 2 or our Fig. 2). Dahm determined different value of the orbital period (4.639094 d) based on the data.

Our analysis shows that measurements of RW Ari in Bookmeyer (1977) were mistakenly mixed with the data of UU Boo (see Fig. 2). Dahm (1992) did not notice the problem and thus he came to meaningless results.

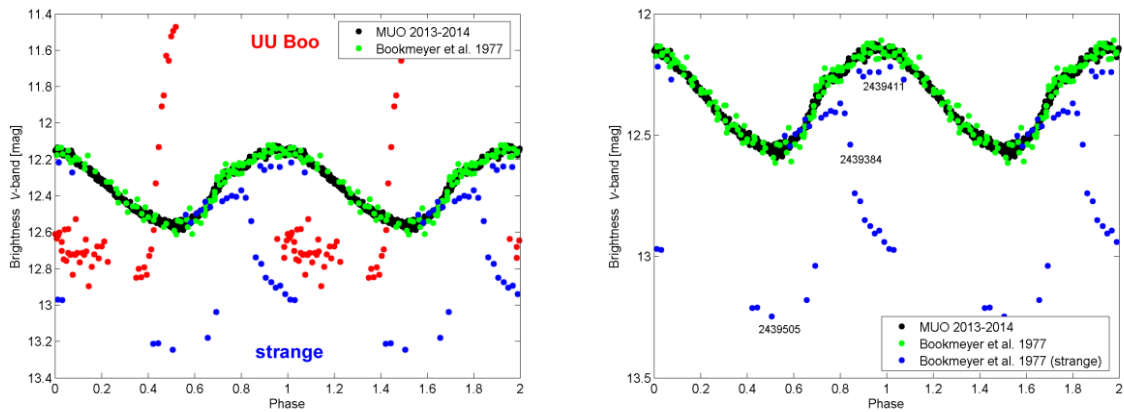

**Figure 2:** The light curve of RW Ari in $V$-band from Bookmeyer et al. (1977) and our observation at MUO phased with the pulsation period, the red part of the curve belongs to UU Boo (the left panel). The light curve is similar like the curve in Dahm (1992). After the selection of data only for RW Ari the light curve is the same as obtained by Wiśniewski (1971).

We reconstructed the variations in RW Ari with subtracted pulsations similarly as Wiśniewski (1971) performed. We found that changes were different in $B$ and $U$ than for $V$-band. In the night JD 2439384 brightness was decreased in all filters (up to 0.6 mag), but in JD 2439505 the star was fainter of about 0.6 mag in $V$-band and in $U$ and $B$-bands RW Ari was even brighter than usually was (see Fig. 3). This strange behaviour indicates some observational problem, not the real changes. Therefore, results from Wiśniewski (1971) should be accepted with attention.

Preliminary analysis of our data (Hájková 2015) brings information about phase variations and possible modulation which was already expected by Edwards (1978). No sign of eclipse with amplitude larger than 0.04 mag was detected.

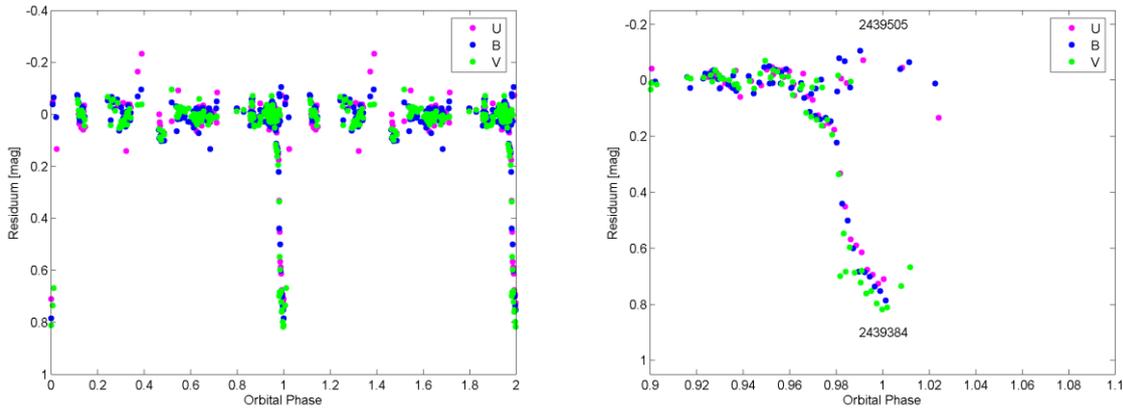

**Figure 3:** The light curve of RW Ari in *UBV*-bands from Bookmeyer et al. (1977) after subtraction of mean pulsation changes and phased with orbital period of 3.1754 d (the left panel). The detail of the primary minimum in the right panel shows completely different behaviour in *U* and *B* filters in the night JD 2439505.

## RZ Cet

Variability of RZ Cet was found by Hoffmeister (1929) on photographic plates. Later Zessewitsch (1932) determined pulsation period as 0.5105 d. Binarity of RZ Cet was discussed by Le Borgne et al. (2007) and Liška et al. (2015) on the basis of variation in *O–C* diagram.

Photometric data from 7 nights for the star is available also in Bookmeyer et al. (1977). Data from one of their nights (JD 2439746) is anomalous as was firstly mentioned in Liška (2015). The brightness of the star is of about 0.4 mag lower in *V*-band than usually is (Fig. 4, the left panel). This could be explained by eclipse. In this case the colour of the star during the eclipse is the same as in the normal nights (Fig. 4, the right panel). There are also some discrepancies in Bookmeyer et al. (1977) and ASAS-3 data in phases 0.7 – 1.0 which could be related to the Blazhko effect. The possible modulation in RZ Cet was already mentioned in Kovács (2005) without its period determination.

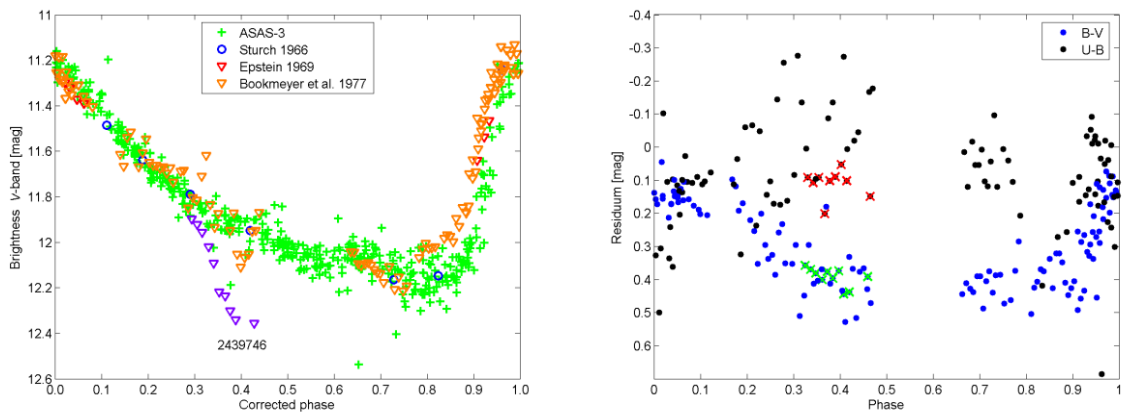

**Figure 4:** The light curve of RZ Cet in *V*-band from Bookmeyer et al. (1977), Sturch (1966), Epstein (1969), and ASAS-3 database phased with pulsation period (the left panel). Variations in colours in observations from Bookmeyer et al. (1977) – green and red crosses mark *B−V* and *U−B* indices from the night JD 2439746 (the right panel).

## Conclusions

We present analysis of VX Her and RW Ari, the historical candidates for EB systems with RR Lyrae component. New photometric data was obtained for both stars. We identified another possible EB system, RZ Cet.

Our study brings some interesting findings. Brightness dips for all three stars were recorded by the same observing group in which Wiśniewski was a member (publications by Fitch et al. 1966, Wiśniewski 1971, Bookmeyer et al. 1977). The photometric measurements were obtained by a single-channel photometer in *UBV* filters in Arizona (Catalina observing stations of LPL). The observing strategy was the same for all three stars – no comparison star was used. Observers used only standard stars from the list Johnson & Harris (1954). Unfortunately, most of the 108 standard stars are included in the International Variable Star Index with note "suspected from variability". For example one extreme variable "standard" star is a Be star V1294 Aql ($\Delta V \sim 1$ mag). For this reason, the brightness drops could be also caused by variability of the used "standard" stars.

The brightness dips were detected in a very short interval from April 1966 (VX Her) to September 1967 (RZ Cet). It is very improbable that all three RR Lyrae stars are EBs, because the group observed only about 200 RR Lyrae stars. High amplitudes of "eclipses" are also highly improbable (VX Her – 0.7 mag, RW Ari – 2 times 0.6 and 0.1 mag, RZ Cet – 0.4 mag). In addition colour variations during "eclipses" in VX Her and RW Ari differ significantly, in RZ Cet the colour is normal. Proper confirmation of eclipses for all mentioned stars was not performed yet. Our observations also contain no sign of eclipses.

We will present a detailed analysis of the VX Her, RW Ari and RZ Cet data in our forthcoming paper(s).

**Acknowledgement**

This research has made use of NASA's Astrophysics Data System and of the International Variable Star Index (VSX) database, operated at AAVSO, Cambridge, Massachusetts, USA. Work on the research have been supported by MUNI/A/1110/2014 and LH14300. MS acknowledges the support of the postdoctoral fellowship programme of the Hungarian Academy of Sciences at the Konkoly Observatory as a host institution. This work was supported by the Brno Observatory and Planetarium.